\documentclass[12pt,a4paper]{article}

\usepackage{amsmath,amssymb}
\usepackage{graphics}
\normalbaselines

\begin{document} 
\bibliographystyle{unsrt}
\vspace*{46mm}

{ \noindent 
 \large \bf  INDUCE REPRESENTATION OF THE POINCARE  \\[6pt] GROUP ON THE LATTICE: SPIN
$1/2$ AND $1$ CASE}
\vspace{36pt}\\
\hspace*{36pt}  Miguel Lorente

\hspace{54pt}  {\it Departamento de F\'{\i}sica, Universidad de Oviedo, 33007 Spain}

\hspace{54pt}   {\it and Institute f\"ur theoretische Physik. Universit\"at T\"ubingen, Germany.} 
\vspace{12pt}\\
\hspace*{36pt}  Peter Kramer

\hspace{54pt}  {\it Institute f\"ur theoretische Physik. Universit\"at T\"ubingen, Germany.}

\vspace{10mm}

\begin{abstract} 

Following standard methods we explore the construction of the discrete Poin\-ca\-r\'e group, the
semidirect product of discrete translations and integral Lorentz transformations, using the
Wigner-Mackey construction restricted to the momentum and position space on the lattice. The orbit
condition, irreducibility and assimptotic limit are discussed.
\end{abstract} 

\section{Introduction}

In a previous paper of this Symposium [14] we have discussed the induced representations of the euclidean
group on the lattice via duality and Fourier transform. In this paper we apply the same method to the
induced representation of Poincar\'e group on the lattice. We explore the problem of irreducibility connected
with the orbit condition, the assymptotic limit of the difference equation for the Klein-Gordon and Dirac
field.

In section 2 we present the realization of the integral Lorentz transformation, that can be factorized
completely with the help of Kac generators, both for the fundamental and spin representation. In section 3 we
introduce several types of Fourier transform that will be used to go from momentum to position space. In
section 4 we reproduce the properties of covariant bispinors in discrete momentum space in similar fashion
to the continuous one. In section 5, we elaborate the induced representation of Poincar\'e group on the
lattice following standard procedure and we compare the results derived from the different types of Fourier
transform.

\section{Integral Lorentz transformations}

A Lorentz transformation is integral if leaves invariant the cubic lattice and at the same
time keeps invariant the bilinear form
\begin{equation}
{x}_{0}^{2}-{x}_{1}^{2}-{x}_{2}^{2}-{x}_{3}^{2}
\end{equation}

According to Coxeter [1] all integral Lorentz transformations (including reflections) are obtained by
combining the operations of permuting the spatial coordinates ${x}_{1},{x}_{2},{x}_{3}$ and changing the
signs of any of the coordinates ${x}_{0},{x}_{1},{x}_{2},{x}_{3}$ together with the operation of adding the
quantity ${x}_{0}-{x}_{1}-{x}_{2}-{x}_{3}$ to each of the four coordinates of a point.

These operations can be described geometrically by the reflections on the plans perpendicular to the vectors
\[{\alpha }_{1}={e}_{1}-{e}_{2},\ {\alpha }_{2}={e}_{2}-{e}_{3},\ {\alpha
}_{3}={e}_{3},\ {\rm
\alpha }_{4}=-\left({{e}_{0}+{e}_{1}+{e}_{2}+{e}_{3}}\right)\]
where $\left\{{{e}_{0},{e}_{1},{e}_{2},{e}_{3},}\right\}$ is an orthonormal basis.

In matritial form these reflections are

\[\begin{array}{ll}
{S}_{1}=\left({\begin{array}{cccc}1&0&0&0\\ 0&0&1&0\\ 0&1&0&0\\ 0&0&0&1\end{array}}\right),\
&{S}_{2}=\left({\begin{array}{cccc}1&0&0&0\\ 0&1&0&0\\ 0&0&0&1\\ 0&0&1&0\end{array}}\right)
\\
 {S}_{3}=\left({\begin{array}{cccc}1&0&0&0\\ 0&1&0&0\\ 0&0&1&0\\
0&0&0&-1\end{array}}\right),\ &{S}_{4}=\left({\begin{array}{cccc}2&1&1&1\\ -1&0&-1&-1\\
-1&-1&0&-1\\ -1&-1&-1&0\end{array}}\right)
\end{array}\]

These reflections constitue a Coxeter group, the Dynkin diagram of which is the following

\bigskip
\begin{center}
\includegraphics{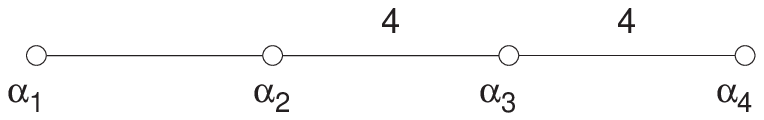}
\end{center}
\medskip
\vskip 13pt

Kac has proved [2] that ${S}_{1},{S}_{2},{S}_{3},{S}_{4}$ generate all the integral Lorentz tranformations
that keep invariant the upper half of the light cone.

This result can be used to factorized any integral Lorentz transformation that belong to the proper
orthochronous group. Let
\begin{equation}
L\equiv \left({\begin{array}{cccc}\mit a&e&f&g\\ b&&&\\ c&&{}^*&\\
d&&&\end{array}}\right)
\end{equation}
be an integral matrix of determinant +1, satisfying
\begin{equation}
Lg{L}^{t}=g\ ,\ g={\rm diag}\left({ 1,-1,-1,-1}\right)
\end{equation}
and also $a\rm \ge  1$. From
\begin{equation}
{a}^{2}-{b}^{2}-{c}^{2}-{d}^{2}=1
\end{equation}
it follows that only one of $b,c,d$ can be zero. Suppose $a>1$. Then we apply
${S}_{1},{S}_{2},{S}_{3}$ to $L$ from the left until $b,c,d$ become non-positive integers. To the resulting
matrix we apply ${S}_{4}$. We get
\[L'\mit =\left({\begin{array}{cccc}a\rm '&\mit e\rm '&\mit f\rm '&\mit g\rm '\\
\mit b\rm '&&&\\
\mit c\rm '&&\mit {}^*&\\ d\rm '&&&\end{array}}\right)\]
with $a\rm '\mit =2a+b+c+d$. Obviously 
\[\left({a+b+c+d}\right)\left({a-b-c-d}\right)=1-2bc-2bd-2cd<0\]
therefore $a+b+c+d<0$ \ or 
\begin{equation}
2a+b+c+d=a'<a
\end{equation}

By iteration of the same algorithm we get
\begin{equation}
a>a'>a''>\ldots >{a}^{\left({k}\right)}\ge 1
\end{equation}

The last inequality is a consequence of the fact that $L$ and ${S}_{4}$ belong to the complete Lorentz
group. Following this process we get an integral matrix with ${a}^{\left({k}\right)}=1$ which is a
combination of ${S}_{1},{S}_{2},{S}_{3}$, giving all the 24 elements of the cubic group on the
lattice.

Therefore a general integral Lorentz transformation of the proper orthochronos type $L$ can be
decomposed as

\begin{equation}
L={P}_{1}^{\eta }{P}_{2}^{\theta }{P}_{3}^{i}{S}_{4}\ldots {S}_{4}{P}_{1}^{\rm
\delta }{P}_{2}^{\varepsilon }{P}_{3}^{\zeta }{S}_{4}\left\{{{S}_{1}^{\alpha
}{S}_{2}^{\beta }{S}_{3}^{\gamma }}\right\}_{\rm all \ permutations}
\end{equation}
where ${P}_{1}={S}_{1}{S}_{2}{S}_{3}{S}_{2}{S}_{1},\ {P}_{2}={S}_{2}{S}_{3}{S}_{2},\
{P}_{3}={S}_{3}$ are the matrix which change sign of $b,c,d$ and $\alpha \mit ,\beta \mit
,\gamma \mit ,\rm
\delta \mit ,\varepsilon \mit ,\zeta \mit ,\eta \mit ,\theta
\mit ,\iota \mit \ldots =0,1$

A particular case of integral Lorentz transformations are the boost or integral transformations of two
inertial systems with relative velocity. The general expresion for these transformations can be obtained
with the help of Cayley parameters [3]. Let us take $n=p=q=0$ and $m,r,s,t,$ integer numbers. We have two
cases, corresponding to two diophantine equations:
\begin{enumerate}
\item[i)] ${m}^{2}-{r}^{2}-{s}^{2}-{t}^{2}=1$
{\footnotesize
\begin{equation}\setlength{\arraycolsep}{0mm}
 L=\left({\begin{array}{cccc}{m}^{2}+{r}^{2}+{s}^{2}+{t}^{2}&2mr&2ms&2mt\\
2mr&{m}^{2}+{r}^{2}-{s}^{2}-{t}^{2}&2rs&2rt\\ 2ms&2rs&{m}^{2}-{r}^{2}+{s}^{2}-{t}^{2}&2st\\
2mt&2rt&2st&{m}^{2}-{r}^{2}-{s}^{2}+{t}^{2}\end{array}}\right)
\end{equation}}

\item[ii)] ${m}^{2}-{r}^{2}-{s}^{2}-{t}^{2}=2$
{\footnotesize
\begin{equation}
L=\left({\begin{array}{cccc}{m}^{2}-1&mr&ms&mt\\ mr&{r}^{2}+1&rs&rt\\
ms&rs&{s}^{2}+1&st\\ mt&rt&st&{t}^{2}+1\end{array}}\right)
\end{equation}}
\end{enumerate}

The solutions of the diophantine equation i) and ii) are obtained by applying all the Coxeter reflections to
the vector $(1,0,0,0)$ in case i) and to the vector $(2,1,1,0)$ in case ii)

From the inspection of (7) if we take the quotient of $L$ with respect the subgroup of all integral
rotations or cubic group we are left with the coset representatives which are not
exhausted by the pure Lorentz transformations (8) or (9), because these are always symmetric
matrices. Therefor we have to add all the integral Lorentz matrix that applied to the vector
$(1,0,0,0)$ gives all the integral vectors of the type $(m,r,s,t)$ with
${m}^{2}-{r}^{2}-{s}^{2}-{t}^{2}=1$. This is equivalent to say that we apply to
$(1,0,0,0)$ not only the matrix $L$ given by (8) but also its square root matrix, namely,

\begin{equation}
\sqrt {L}=\left({\begin{array}{cccc}m&r&s&t\\
r&1+{\frac{{r}^{2}}{m+1}}&{\frac{rs}{m+1}}&{\frac{rt}{m+1}}\\
s&{\frac{rs}{m+1}}&1+{\frac{{s}^{2}}{m+1}}&{\frac{st}{m+1}}\\
t&{\frac{rt}{m+1}}&{\frac{st}{m+1}}&1+{\frac{{t}^{2}}{m+1}}\end{array}}\right)
\end{equation}

In position space the space-time coordinates of the lattice ${x}_{\mu }$ are integer numbers. They
transform under integral Lorentz transformations into integral coordinates. The same is true for the
increments ${\Delta \mit x}_{\mu }$.

In momentum space the components of the four-momentum are not integer numbers but they can be constructed
with the help of integral coordinates, namely,

\[p_{\mu }\equiv {\mit m}_{\mit 0}\mit c\left({{\frac{c\Delta \mit
t}{{\left({{\left({c\Delta \mit t}\right)}^{2}-{\left({\Delta \vec{\mit
x}}\right)}^{2}}\right)}^{1/2}}},{\frac{\Delta \vec{\mit x}}{{\left({{\left({\mit c\Delta
\mit t}\right)}^{\mit 2}\mit -{\left({\Delta \vec{\mit x}}\right)}^{2}}\right)}^{\mit
1/2}}}}\right)\]

If ${\Delta \mit x}_{\mu }$ transform under integral Lorentz transformations the new ${p'}_{\mu }$ will be
given in terms of integral ${\Delta \mit x'}_{\mu }$.

Using the homomorphism between the groups $SO(3,1)$ and $SL(2,C\hspace{-9pt}/)$ we obtain the
representation of integral Lorentz transformations in 2-dimensional complex matrices. From  the
knowledge of the Cayley parameters [5] we read off the matrix elements of $\alpha \in SL(2,C\hspace{-9pt}/)$

\begin{equation}
\alpha \mit ={\frac{1}{\sqrt {\Delta }}}\left({\begin{array}{cc}m+t+i\left({n-\rm
\lambda }\right),&-p+r+i\left({q+s}\right)\\ p+r+i\left({q-s}\right),&m-t-i\left({n+\lambda
}\right)\end{array}}\right)
\end{equation}

\begin{equation}
\Delta \mit =\ \det\alpha \mit ={m}^{2}-{r}^{2}-{s}^{2}-{t}^{2}+{m}^{2}+{p}^{2}+{q}^{2}-{
\lambda }^{2}
\end{equation}

For instance, we calculate the 2-dimensional representation of the Coxeter reflection ${S}_{i}$ multiplied
by the parity operator $P$ in order  to get an element of the proper Lorentz group) identifying its matrix
elements with Lorentz matrix written in terms of Cayley parameters: Easy calculations give the unique
solutions:

\begin{eqnarray}
D\left({P{S}_{1}}\right)&=&\rm \pm {\frac{1}{\sqrt {2}}}\left({\begin{array}{cc}0&-1-i\\
1-i&0\end{array}}\right) \\
D\left({P{S}_{2}}\right)&=&\rm \pm {\frac{1}{\sqrt {2}}}\left({\begin{array}{cc}i&1\\
-1&-i\end{array}}\right) \\
D\left({P{S}_{3}}\right)&=&\rm \pm \left({\begin{array}{cc}i&\\ &-i\end{array}}\right) \\
D\left({P{S}_{4}}\right)&=&\rm \pm {\frac{1}{\sqrt {2}}}\left({\begin{array}{cc}0&1-i\\
-1-i&2i\end{array}}\right)
\end{eqnarray}

The integral Lorentz transformations without rotations as given in (8) and (9) have a 2-dimensional
representation making $n=p=q=\lambda  =0$ in (11) and the choice $\Delta 
={m}^{2}-{r}^{2}-{s}^{2}-{t}^{2}=1\ \rm or\ 2$ in (12).

In order to complete the picture we have to add the 2-dimensional representation of the matrix $\sqrt {L}$
given in (10) which turns out to be

\begin{equation}
\alpha \mit ={\frac{1}{\sqrt {2\left({m+1}\right)}}}\left({\begin{array}{cc}\rm
m+1+t&r-is\\ r+is&m+1-t\end{array}}\right)
\end{equation}

We give also the $2\times  2$ matrix representation of the discrete momentum. Let
${p}_{\mu }$ a four momentum which is obtained by applying all the integral Lorentz
transformations, given by (7) divided by the cubic group, to the vector
$\left({{m}_{0}c,0,0,0}\right)$. In 2-dimensional matrix form this is equivalent to apply the
matrix $\alpha$ given by (17) to the unit matrix multiply by ${m}_{0}c$:

\begin{equation}
\alpha \left({\begin{array}{cc}{\mit m}_{\mit 0}\mit c&0\\
0&{m}_{0}c\end{array}}\right){\alpha }^{\mit +}\mit
={m}_{0}c\left({\begin{array}{cc}m+t&r-is\\ r+is&m-t\end{array}}\right)
\end{equation}
which is of the standard form if we identify the components of the 4-momentum as

\begin{equation}
{p}_{\mu }={m}_{0}c\left({m,r,s,t}\right)
\end{equation}
with $m,r,s,t$ integral numbers, satisfying
${m}^{2}-{r}^{2}-{s}^{2}-{t}^{2}=1$.

\section{Fourier transform on the lattice}

In order to go from position space to momentum space on the lattice we can define several restrictions of
the continuous variables of Fourier transform to the discrete variables on the lattice.

\vspace*{12pt}\noindent{\bf I type. Discrete position and momentum variables of finite rank}\vspace*{12pt} 

We construct an orthonormal basis [6]

\begin{equation}
{f}_{j}\left({{p}_{m}}\right)={\left({{\frac{1+{\frac{1}{2}}i\varepsilon {\mit p}_{\mit
m}}{\mit 1-{\frac{1}{2}}\varepsilon {\mit p}_{\mit m}}}}\right)}^{j},\ j=0,1,\ldots N-1
\end{equation}

\[{p}_{m}={\frac{2}{\varepsilon }}tg{\frac{\pi }{\mit N}}m,\ \ m=0,1,\ldots N-1\]
that satisfy periodic boundary conditions:

\[{f}_{0}\left({{p}_{m}}\right)={f}_{N}\left({{p}_{m}}\right),\ \ {p}_{m+N}={p}_{m},\]
orthogonality relations:

\begin{equation}
{\frac{1}{N}}\sum\limits_{\mit j\ =\ 0}^{\mit N-1} {\mit f}_{\mit j}^{\mit
{}^*}\left({{\mit p}_{\mit m}}\right){\mit f}_{\mit j}\left({{\mit p}_{\mit m\rm
'}}\right)\mit ={\delta }_{mm'}
\end{equation}
and completness relations:

\begin{equation}
{\frac{1}{N\varepsilon }}\sum\limits_{\mit m\ =\ 0}^{\mit N-1} {\mit f}_{\mit
j}^{\mit {}^*}\left({{\mit p}_{\mit m}}\right){\mit f}_{\mit j'}\left({{\mit p}_{\mit m}}\right)\mit
={\frac{1}{\varepsilon }}{\delta }_{jj'}
\end{equation}

The finite Fourier transform reads

\begin{equation}
{\hat{F}}_{m}={\frac{1}{\sqrt {N}}}\sum\limits_{\mit j\ =\ 0}^{\mit N-1} {\mit f}_{\mit
j}^{\mit {}^*}\left({{\mit p}_{\mit m}}\right){\mit F}_{\mit j}
\end{equation}
for some periodic function ${F}_{j}$ on the lattice

\[{F}_{j+N}={F}_{j}\]

If we write ${f}_{j}\left({{p}_{m}}\right)\equiv \exp\left({\mit i{\frac{2\rm
\pi }{\mit N}}mj}\right)$ this transform coincide with standard finite Fourier transform [7].

\vspace*{12pt}\noindent{\bf II type. Discrete position and continuous momentum}\vspace*{12pt}

When we restrict the position variables in continuous Fourier transform to discrete values we obtain the
Fourier series.

\begin{eqnarray}
& &{\rm Orthonormal\  basis:\ }{\left\{{{\frac{1}{\sqrt {2\pi }}}{e}^{ikj\rm
\varepsilon }}\right\}}_{j=-\infty }^{\infty }\equiv {\mit f}_{\mit j}\left({\mit
k}\right) \\
& &{\rm Orthogonality \  relations:\ } 
\int_{-\pi /\varepsilon }^{\pi /\varepsilon }{f}_{j}^{{}^*}\left({k}\right){f}_{j\rm
'}\left({k}\right)dk={\frac{1}{\varepsilon }}{\delta }_{jj'\mit \ } \\
& &{\rm Completness \  relation:\ } 
\sum\limits_{\mit j=-\infty }^{\mit \infty } {\mit f}_{\mit j}^{\mit {}^*}\left({\mit
k}\right){\mit f}_{\mit j}\left({\mit k'}\right)\mit ={\frac{1}{\varepsilon }}\delta
\left({\mit k-k'}\right)
\end{eqnarray}

Fourier expansion: for a periodic function $F\left({k}\right)$

\begin{equation}
F\left({k}\right)=\sum\limits_{\mit j=-\infty }^{\mit \infty } {\mit f}_{\mit j}^{\mit
{}^*}{\mit F}_{\mit j}
\end{equation}

\begin{equation}
{F}_{j}=\int_{-\pi /\mit a}^{\pi /\mit a}{f}_{j}\left({k}\right)F'\left({\mit
k}\right)\mit dk
\end{equation}

Now we make the change of variable
\begin{equation}
P={\frac{2}{\varepsilon }}tg\ {\frac{1}{2}}k\varepsilon
\end{equation}

\begin{equation}
{f}_{j}\left({ p}\right)={\frac{1}{\sqrt {2\pi
}}}{\left({{\frac{1+{\frac{1}{2}}i\varepsilon p}{1-{\frac{1}{2}}i\varepsilon p}}}\right)}^{j}
\end{equation}
and the orthogonality relations become
\begin{equation}
\int_{-\infty }^{\infty }{f}_{j}^{{}^*}\left({
p}\right){f}_{j'}\left({ p}\right){\frac{dp}{1+{\frac{1}{4}}{\varepsilon
}^{2}{p}^{2}}}={\frac{1}{\varepsilon }}{\delta }_{jj'}
\end{equation}
and the completness relation
\begin{equation}
\sum\limits_{j=-\infty }^{\infty }
{f}_{j}^{{}^*}\left({p}\right){f}_{j}\left({p'}\right)=\left({1+{\frac{1}{4}}{\varepsilon
}^{2}{p}^{2}}\right)\delta \left({p-p'}\right)
\end{equation}

Notice that ${f}_{j}\left({p}\right)$ may not be a periodic function.

\vspace*{12pt}\noindent{\bf III type: discrete position and discrete momentum of infinite rank}\vspace*{12pt}

We construct an orthonormal basis
\begin{equation}
{f}_{n}\left({j}\right)={\left({{\frac{1+{\frac{1}{2}}i\varepsilon
k}{1-{\frac{1}{2}}i\varepsilon p}}}\right)}^{n},\ \ k,n\in {\mathcal{Z}}
\end{equation}
satisfying
\begin{equation}
\lim\limits_{N\rightarrow \infty }^{}\ {\frac{1}{2N+1}}\sum\limits_{\mit
n=-N}^{\mit N} {f}_{n}\left({k}\right){f}_{n}\left({k'}\right)={\delta }_{kk'}
\end{equation}

\noindent (Proof: For $k\ne k'$ we use the identity 
\vspace{-3mm}$$\rm 1+\cos\theta +\cos2\theta +...+cosN\theta ={\frac{1}{2}}+{\frac{\sin\
\left({N+{\frac{1}{2}}}\right)\theta }{\sin\ \left({{\frac{\theta }{2}}}\right)}}$$
with ${e}^{i\theta }\equiv {f}_{1}^{{}^*}\left({k}\right){f}_{1}\left({k'}\right)$).

\bigskip The completness relation is now:
\begin{equation}
\sum\limits_{k=-\infty }^{\infty }
{f}_{n}^{{}^*}\left({k}\right){f}_{n'}\left({k}\right)=\lim\limits_{L\rightarrow \infty }^{}\
{\delta }_{L}\left({n-n'}\right)
\end{equation}
where
\begin{equation}
{\delta }_{L}\left({n-n'}\right) =\sum\limits_{j=-L}^{L}
{f}_{n}\left({k}\right){f}_{n'}^{{}^*}\left({k}\right)
\end{equation}
is a $\delta$ sequence satisfying

\begin{equation}
\lim\limits_{N\rightarrow \infty }^{}\ {\frac{1}{2N+1}}\sum\limits_{n=-N}^{N} {\delta
}_{L}\left({n-n'}\right)=1
\end{equation}

\noindent (Proof:\begin{eqnarray*}
& &{\frac{1}{2N+1}} \sum\limits_{n=-N}^{N}
{\delta}_{L}\left({n-n'}\right)=\\ {\smallskip}
&=&  1+\sum\limits_{k=1}^{L}
{\frac{1}{2N+1}}\sum\limits_{n=-N}^{N}
\left({{f}_{n}\left({k}\right){f}_{n'}^{{}^*}\left({k}\right)+c.c.}\right) =\\
{\smallskip}
&=&1+\sum\limits_{k=1}^{L} {\frac{1}{2N+1}}\left({{\frac{\sin\
\left({N+{\frac{1}{2}}}\right)\theta }{\sin\ {\frac{1}{2}}\theta
}}{f}_{n'}^{*}\left({k}\right)+c.c.}\right) 
{\longrightarrow 1 \atop {\scriptstyle N \rightarrow \;\infty}}
\end{eqnarray*}
for all $L$, as required)

The Fourier transform becomes:

\begin{equation}
\hat{ F}\left({ k}\right) =\lim\limits_{N\rightarrow \infty }^{}\
{\frac{1}{2N+1}}\sum\limits_{n=-N}^{N} {f}_{n}\left({k}\right){F}_{n}
\end{equation}
where ${ F}_{ n}\rightarrow 0$ when $n\rightarrow \infty$
and $\frac{1}{2N+1} \sum\limits_{n=-N}^{N} {F}_{n}\ {\longrightarrow \; 0 \atop
{\scriptscriptstyle N \rightarrow \;\infty}}$

\begin{equation}
{ F}_{ n} =\sum\limits_{k=-\infty }^{\infty }
{f}_{n}\left({k}\right)\hat{F}\left({k}\right)
\end{equation}

The Fourier transform of type III was introduced in [8]. When

$$ n,n'\rightarrow \infty ,\ \ \varepsilon \rightarrow 0\ \ \ m\varepsilon \rightarrow x$$
$${f}_{n}\left({k}\right) \rightarrow {e}^{ikx}$$
the orthogonality relations converges

$$ \lim\limits_{N\rightarrow \infty }^{}\
{\frac{1}{2N+1}}\sum\limits_{n=-N}^{N}
{f}_{n}^{{}^*}\left({k}\right){f}_{n}\left({k'}\right)=\int_{-\pi }^{\pi }{e}^{-ikx}{e}^{ik'x}dk$$
the completness relations becomes

$$\sum\limits_{j=-\infty }^{\infty }
{f}_{n}^{{}^*}\left({k}\right){f}_{n'}\left({k}\right)\rightarrow \sum\limits_{k=-\infty }^{\infty
} {e}^{-ikx}{e}^{ikx'}$$
and the Fourier transform converges to:

\begin{eqnarray*}
\hat{ F}\left({ k}\right) &=&\int_{}^{}{e}^{ixk}F\left({x}\right)dx \\
F\left({k}\right)&=&\sum\limits_{k=-\infty }^{\infty } {e}^{ikx}\hat{F}\left({k}\right)
\end{eqnarray*}

\section{Dirac and vector representation of the Lorentz group and covariant wave equations}

Let $ L\left({\alpha }\right)$ be an element of the proper Lorentz group corresponding to the
element $\alpha \in SL\left({2,C}\right)$ and $I_s$ the parity operator. If the components of
four-momentum are written as $2\times 2$ matrix

\begin{equation}
\tilde{p}\equiv {p}^{\mu }{\sigma }_{\mu }={p}^{0}{\sigma
}_{0}+{p}^{i}{\sigma }_{i}
\end{equation}
where ${\sigma}_0=1$ and ${\sigma}_i$ are the Pauli matrices.

The transformations of $\tilde{\rm p}$ under parity and $SL(2,C)$ are

\[{I}_{s}:\tilde{p}\rightarrow {\tilde{p}}^{s}={p}^{0}{\sigma }_{0}-{p}^{i}{\sigma
}_{i}=\left({\det\ \tilde{p}}\right){\left({\tilde{p}}\right)}^{-1}\]

\[\alpha :\tilde{p}\rightarrow \alpha \tilde{p}{\alpha }^{+}\]

\begin{equation}
{\tilde{p}}^{s}\rightarrow {\left({{\alpha
}^{+}}\right)}^{-1}{\tilde{p}}^{s}{\alpha }^{-1}
\end{equation}

It follows
\begin{equation}
I_sL(\alpha)I_s^{-1} = L\left((\alpha^+)^{-1}\right)
\end{equation}

Therefore the matrix ${\left({{a}^{+}}\right)}^{-1}$ gives an other 2-dimensional
representation of the Lorentz group non-equivalent to $\alpha \in SL\left({2,C}\right)$.
In order to enlarge the proper Lorentz group with space reflection we take both representations
$\alpha$ and ${\left({{\alpha }^{+}}\right)}^{-1}$.

Let $\pi \equiv \left\{{\mit I,{I}_{s}}\right\}$ the space reflection group and $\alpha
\in SL\left({2,C}\right)$, then the semidirect product 

$$SL\left({2,C}\right)\otimes \pi$$
with the multiplication law

\begin{eqnarray}
\left({\alpha ,\pi }\right)\ \left({\alpha ',\pi '}\right)&=&\left({\alpha
\alpha ',\pi \pi '}\right)\ if\ \pi =I \\
\left({\alpha ,\pi }\right)\ \left({\alpha ',\pi '}\right)&=&\left({\alpha
{({{\alpha '}^{+}})}^{-1},\pi \pi '}\right)\ if\ \pi ={I}_{s}
\end{eqnarray}
form a group.

This group has a 4-dimensional representation, a particular elements of which is

\begin{equation}
\overline{D}\left({\alpha ,I}\right)=\left({\begin{array}{cc}\alpha
&0\\ 0&{\left({{\alpha }^{+}}\right)}^{-1}\end{array}}\right),\
\overline{D}\left({e,{I}_{s}}\right)=\left({\begin{array}{cc}0&{\sigma }_{0}\\ {\sigma
}_{0}&0\end{array}}\right)
\end{equation}
that satisfy

\begin{equation}
\overline{D}\left({e,{I}_{s}}\right)\overline{D}\left({\alpha
,I}\right)\overline{D}\left({e,{I}_{s}^{-1}}\right)=,\
\overline{D}\left({{\left({{\alpha }^{+}}\right)}^{-1},I}\right)
\end{equation}

With respect to this representation a-4-component spinor in momentum space $\psi
\left({p}\right)$ transform as follows

\begin{eqnarray}
U\left({\alpha ,I}\right)\overline{\psi
}\left({p}\right)&=&\overline{D}\left({\alpha ,I}\right)\overline{\psi }\left({{L}^{-1}\left({\alpha
}\right)p}\right) \\
U\left({e,{I}_{s}}\right)\overline{\psi
}\left({p}\right)&=&\overline{D}\left({e,{I}_{s}}\right)\overline{\psi }\left({{I}_{s},p}\right)
\end{eqnarray}

Using a similarity transformation we obtain an equivalent representation

\[D\left({\alpha ,\pi }\right)=M\overline{D}\left({\alpha ,\pi }\right){M}^{-1}\]
with
\[M={\frac{1}{\sqrt {2}}}\left({\begin{array}{cc}{\sigma }_{0}&{\sigma }_{0}\\
{-\sigma }_{0}&{\sigma }_{0}\end{array}}\right)\]

In this representation

\begin{eqnarray}
D\left({\alpha ,I}\right)&=&{\frac{1}{2}}\left({\begin{array}{cc}{\alpha +\left({{\alpha
}^{+}}\right)}^{-1},&{\mit -\alpha +\left({{\alpha }^{+}}\right)}^{-1}\\ {\mit -\alpha
+\left({{\alpha }^{+}}\right)}^{-1}\mit ,&{\alpha +\left({{\alpha }^{+}}\right)}^{\rm
-1}\end{array}}\right) \\
D\left({e,{I}_{s}}\right)&=&\left({\begin{array}{cc}{\sigma }_{0}&0\\
0&{\mit -\sigma }_{0}\end{array}}\right)
\end{eqnarray}

The new four-spinor

\[\psi \left({p}\right)=M\overline{\psi }\left({p}\right)\]

transform as

\begin{eqnarray}
U\left({\alpha ,I}\right)\psi \left({p}\right)&=&D\left({\alpha ,I}\right)\psi
\left({{L}^{-1}\left({\alpha }\right)p}\right) \\
U\left({e,{I}_{s}}\right)\psi \left({p}\right)&=&D\left({e,{I}_{s}}\right)\psi
\left({{I}_{s},p}\right)
\end{eqnarray}

The Dirac wave equation can be considered as a consequence of the relativistic invariance [9]

In the rest system we want a projection operator that selects one irreducible representation out
of the Dirac representatition. This is achieve in the rest system by

\begin{equation}
Q={\frac{1}{2}}\left({I+\beta }\right),\ \ \ \beta \equiv \left({\begin{array}{cc}{\sigma
}_{0}&0\\ 0&-{\sigma }_{0}\end{array}}\right)
\end{equation}

In order to get the projection operator in an arbitrary system we apply $D(\kappa)$ given by (49) and (17)

\begin{equation}
Q\rightarrow
Q\left({\kappa}\right)={D}^{-1}\left({\kappa}\right)QD\left({\kappa}\right)={\frac{1}{2}}\left({I+W\left({\kappa}\right)}\right)
\end{equation}
where 
\begin{equation}
W\left({\kappa}\right)={\frac{1}{2}}\left({\begin{array}{cc}{\kappa}^{+}\kappa+{\left({{\kappa}^{+}\kappa}\right)}^{-1},&-{\kappa}^{+}\kappa+{\left({{\kappa}^{+}\kappa}\right)}^{-1}\\
{\kappa}^{+}\kappa-{\left({{\kappa}^{+}\kappa}\right)}^{-1},&-{\kappa}^{+}\kappa-{\left({{\kappa}^{+}\kappa}\right)}^{-1}\end{array}}\right)
\end{equation}

Using the identities

 \parbox{11cm}{ 
\begin{eqnarray*}
& &{\left({{\kappa}^{+}\kappa}\right)}^{-1}={\frac{1}{{m}_{0}c}}\sum\limits_{\mu =0}^{3}
{\sigma }^{\mu }{p}_{\mu } \\
& &{\kappa}^{+}\kappa={\frac{1}{{m}_{0}c}}{\sigma }^{0}{p}_{0}-\sum\limits_{j=1}^{3} {\sigma
}^{j}{p}_{j}
\end{eqnarray*}}\hfill
\parbox{1cm}{\begin{eqnarray}\end{eqnarray}}

\noindent we find
\begin{equation}
W\left({\kappa}\right)={\frac{1}{{m}_{0}c}}\sum\limits_{\mu =0}^{3} {\gamma }^{\mu
}{p}_{\mu }
\end{equation}
where ${\gamma }^{\mu}$ are Dirac matrices with the realization

\[{\gamma }^{0}=\left({\begin{array}{cc}{\sigma }^{0}&0\\
0&-{\sigma }^{0}\end{array}}\right)\ \ \ ,\ \ \ {\gamma }^{\varphi
}=\left({\begin{array}{cc}0&{\sigma }^{j}\\ -{\sigma }^{j}&0\end{array}}\right)\]

\[{\gamma }^{0}={\gamma }_{0}\ \ \ ,\ \ \ {\gamma }^{\varphi }={-\gamma }_{j}\]

Collecting these result we obtain the Dirac equation in momentum space

\[Q\left({\kappa}\right)\psi \left({p}\right)={\frac{1}{2}}\left({I+W\left({\kappa}\right)\psi
\left({p}\right)}\right)=\psi \left({p}\right)\]
or

\begin{equation}
\sum\limits_{\mu =0}^{3} \left({{\gamma }^{\mu }{p}_{\mu }-{m}_{0}cI}\right)\psi
\left({p}\right)=0
\end{equation}

(An equivalent method can be used applying to the projection operator the Foldy-Wouthuysen
transformation [10])

We apply the operator $\sum\left({{\gamma
}^{\mu }{p}_{\mu }-{m}_{0}c}\right)$ from the left to (59) and obtain 

\begin{equation}
\left({{p}^{\mu }{p}_{\mu }-{m}_{0}^{2}{c}^{2}}\right)\psi
\left({p}\right)=0
\end{equation}

The Dirac equation is invariant under the group $\left({\alpha ,\pi }\right)$ defined before.
In other words, if $\psi \left({p}\right)$ is a solution of the Dirac equation, so is $\rm
U\left({\alpha ,\pi }\right)\psi \left({p}\right)$. Put $\pi=I$. Then

$$U\left({\alpha ,I}\right)\psi \left({p}\right)=D\left({\alpha ,I}\right)\psi
\left({{L}^{-1}p}\right)$$

$${D}^{-1}\left({\alpha }\right)Q\left({\kappa}\right)D\left({\alpha
}\right)=Q\left({x\alpha }\right)$$

$$Q\left({\kappa \alpha }\right)\psi \left({{L}^{-1}p}\right)=\psi \left({{L}^{-1}p}\right)$$

Then

\begin{eqnarray*}
Q\left({\kappa}\right)U\left({\alpha ,I}\right)\psi
\left({p}\right)&=&Q\left({\kappa}\right)D\left({\alpha ,I}\right)\psi \left({{L}^{-1}\left({\alpha
}\right)p}\right)= 
D\left({\alpha ,I}\right)Q\left({\kappa \alpha }\right)\psi \left({{L}^{-1}\left({\alpha
}\right)p}\right)=\\
&=& D\left({\alpha ,I}\right)\psi \left({{L}^{-1}\left({\alpha
}\right)p}\right)=U\left({\alpha ,I}\right)\psi \left({p}\right)
\end{eqnarray*}
as required. For the space reflection

$$Q\left({{I}_{s}\kappa}\right)\psi \left({{I}_{s}p}\right)=\psi \left({{I}_{s}p}\right)$$

$${D}^{-1}\left({{I}_{s}}\right)\rm
Q\left({\kappa}\right)D\left({{I}_{s}}\right)=Q\left({{I}_{s}p}\right)$$

$$U\left({e,{I}_{s}}\right)\psi \left({p}\right)=D\left({{I}_{s}}\right)\psi
\left({{I}_{s}p}\right)$$
we get

\begin{eqnarray*}
Q\left({\kappa}\right)U\left({e,{I}_{s}}\right)\psi
\left({p}\right)&=&Q\left({\kappa}\right)D\left({{I}_{s}}\right)\psi
\left({{I}_{s}p}\right)=D\left({{I}_{s}}\right)Q\left({{I}_{s}\kappa}\right)\psi
\left({{I}_{s}p}\right)= \\
&=& D\left({{I}_{s}}\right)\psi
\left({{I}_{s}p}\right)=U\left({e,{I}_{s}}\right)\psi \left({p}\right)
\end{eqnarray*}
as required. Notice that all properties of Dirac representation in continuous momentum space are
carried out to the discrete momentum space without modification.

For the vector representation of the Lorentz group we take an element $L(p)$ such that takes the
momentum in the rest system to an arbitrary system.

Intherest system the projection operator is

\begin{equation}
Q={\frac{1}{2}}\left({1-g}\right)=\left({\begin{array}{cccc}0&&&\\
&1&&\\
&&1&\\
&&&1\end{array}}\right)
\end{equation}

Then 

\begin{equation}
{L}^{-1}\left({p}\right)QL\left({p}\right)\equiv
Q\left({p}\right)={g}_{\mu \nu }-{\frac{{p}_{\mu }{p}_{\nu }}{{m}_{0}^{2}}}
\end{equation}

And the wave equation in momentum space becomes $Q\left({p}\right)\psi \left({p}\right)=\psi
\left({p}\right)$ 
or 

\begin{equation}
\left({{g}_{\mu \nu }-{\frac{{p}_{\mu }{p}_{\nu
}}{{m}^{2}}}}\right){\psi }_{\mu }\left({p}\right)={\psi }_{\nu }\left({p}\right)
\end{equation}
where ${\psi }_{\mu }\left({p}\right)$ is a 4-component vector function.

In order to construct wave equation in configuration space for the Dirac and vector
representation we use the Fourier transform given in section 3. We take the three types

\vspace{12pt}\noindent {\bf Type I}. Boundary condition imposes finite values for the momentum $p_{\mu}$

\begin{equation}
{p}_{\mu }={\frac{2}{{\varepsilon }_{\mu }}}tg\ {\frac{\pi
}{N}}{m}_{\mu }\ \ \ {m}_{\mu }=0,1,...,\ N-1
\end{equation}

Define the difference operator [8]

\begin{equation}
{\delta }_{\mu }^{+}={\frac{1}{{\varepsilon }_{\mu }}}{\Delta
}_{\mu }\prod\limits_{\nu \ne \mu }^{} {\tilde{\Delta }}_{\nu },\ {\delta }_{\mu
}^{-}={\frac{1}{{\varepsilon }_{\mu }}}{\nabla }_{\mu }\prod\limits_{\nu \ne \mu }^{}
{\tilde{\nabla }}_{\nu },\ \mu ,\nu =1,2,3
\end{equation}

\begin{equation}
{\eta }^{+}=\prod\limits_{\mu =0}^{3} {\tilde{\Delta }}_{\mu }\ \ \ ,\
\ \ {\eta }^{-}=\prod\limits_{\mu =0}^{3} {\tilde{\nabla }}_{\mu }
\end{equation}

Multiply the Dirac equation (60) by

$${\eta }^{+}f\left({{n}_{\mu },{p}_{\mu }}\right)$$
with 

\begin{equation}
f\left({{n}_{\mu },{p}_{\mu }}\right)=\prod\limits_{\mu =0}^{3}
{\left({{\frac{1+{\frac{i}{2}}{\varepsilon }_{\mu }{p}_{\mu }}{1-{\frac{i}{2}}{\varepsilon }_{\mu
}{p}_{\mu }}}}\right)}^{{n}_{\mu }}\ \ {m}_{\mu },{n}_{\mu }\in {\mathcal{Z}}
\end{equation}
the plane waves on the lattice, and using the properties of theses functions

\begin{equation}
{\frac{1}{{\varepsilon }_{\mu }}}{ \Delta }_{ \mu }f\left({{n}_{\mu },{p}_{\mu
}}\right)=i{p}_{\mu }{\tilde{\Delta }}_{\mu }f\left({{n}_{\mu },{p}_{\mu
}}\right)
\end{equation}
we obtain

$$\left({i{\gamma }^{\mu }{\delta }_{\mu }^{+}-{m}_{0}c{\eta }^{+}}\right)\psi
\left({{p}_{\mu }}\right)f\left({{n}_{\mu },{p}_{\mu }}\right)=0$$

Adding for $m=0,1\ldots N-1$, we get the inverse Fourier transform

\begin{equation}
\psi \left({{\eta }_{\mu }}\right)=\sum\limits_{m=0}^{N-1} \psi \left({{p}_{\mu
}}\right)f\left({{n}_{\mu },{p}_{\mu }}\right)
\end{equation}
that satisfies the Dirac equation on the lattice

\begin{equation}
\left({i{\gamma }^{\mu }{\delta }_{\mu }^{+}-{m}_{0}c{\eta }^{+}}\right)\psi
\left({{n}_{\mu }}\right)=0
\end{equation}

\vspace{12pt}\noindent {\bf Type II}. The momentum is continuous and has the range $-\infty <p<\infty$

The plane waves appearing in the Fourier transform are

\begin{equation}
f\left({{n}_{\mu },{p}_{\mu }}\right)={\frac{1}{\sqrt {2\pi
}}}\prod\limits_{\mu =0}^{3} {\left({{\frac{1+{\frac{1}{2}}i{\varepsilon }_{\mu }{p}_{\mu
}}{1-{\frac{1}{2}}i{\varepsilon }_{\mu }{p}_{\mu }}}}\right)}^{{n}_{\mu }}
\end{equation}

As before we multiply the Dirac equation in momentum space by

$${\eta }^{+}f\left({{n}_{\mu },{p}_{\mu }}\right)$$
and using the property

\begin{equation}
{\frac{1}{{\varepsilon }_{\mu }}}{\Delta }_{\mu }f\left({{n}_{\mu },{p}_{\mu
}}\right)=i{p}_{\mu }{\tilde{\Delta }}_{\mu }f\left({{n}_{\mu },{p}_{\mu }}\right)
\end{equation}
we get 

\begin{equation}
\left({i{\gamma }^{\mu }{\delta }_{\mu }^{+}-{m}_{0}c{\eta }^{+}}\right)\rm
\psi
\left({{p}_{\mu }}\right)f\left({{n}_{\mu },{p}_{\mu }}\right)=0
\end{equation}

Integrating for $p$ and puting

\begin{equation}
\psi \left({{n}_{\mu }}\right)=\int_{-\infty }^{\infty }\psi \left({{p}_{\mu
}}\right)f{}^*\left({{n}_{\mu },{p}_{\mu }}\right)\prod\limits_{\mu =0}^{3}
\left({{\frac{d{p}_{\mu }}{1+{\frac{1}{4}}{\varepsilon }_{\mu }^{2}{p}_{\mu
}^{2}}}}\right)
\end{equation}
we obtain the desired Dirac equation in configuration space.

\vspace{12pt}\noindent {\bf Type III}. The momentum is discrete and was given in section 2, namely

\begin{equation}
{k}_{\mu }={m}_{0}c\left({m,r,s,t}\right)\ \ \ ,\ \ \
{m}^{2}-{r}^{2}-{s}^{2}-{t}^{2}=1\ \ \ ,\ \ \ {m,r,s,t} \in {\mathcal{Z}}
\end{equation}

The plane waves are

\begin{equation}
f\left({{n}_{\mu },{k}_{\mu }}\right)=\prod\limits_{\mu =0}^{3}
{\left({{\frac{1+{\frac{1}{2}}i{\varepsilon }_{\mu }{k}_{\mu }}{1-{\frac{1}{2}}i{\varepsilon
}_{\mu }{k}_{\mu }}}}\right)}^{{n}_{\mu }}
\end{equation}

Notice that we do not impose boundary conditions therefore the parameters $k_{\mu}$ are discrete
and infinite. We multiply the Dirac equation in momentum space by

$${\eta }^{+}f\left({{n}_{\mu },{k}_{\mu }}\right)$$
and using

\begin{equation}
{\frac{1}{{\varepsilon }_{\mu }}}{\Delta }_{\mu }f\left({{n}_{\mu },{k}_{\mu
}}\right)=i{k}_{\mu }{\tilde{\Delta }}_{\mu }f\left({{n}_{\mu },{k}_{\mu }}\right)
\end{equation}
we get as before

\begin{equation}
\left({i{\gamma }^{\mu }{\delta }_{\mu }^{+}-{m}_{0}c{\eta }^{+}}\right)\rm
\psi
\left({{k}_{\mu }}\right)f\left({{n}_{\mu },{k}_{\mu }}\right)=0.
\end{equation}

Summing over all values of Cayley parameter that define a four-momentum we get

\begin{equation}
\left({i{\gamma }^{\mu }{\delta }_{\mu }^{+}-{m}_{0}c{\eta }^{+}}\right)\rm
\psi
\left({{n}_{\mu }}\right)=0
\end{equation}
where

\begin{equation}
\psi \left({{n}_{\mu }}\right)=\sum\limits_{m,r,s,t}^{} \psi \left({{k}_{\mu
}}\right)f\left({{n}_{\mu },{k}_{\mu }}\right).
\end{equation}

Notice that $\psi \left({{k}_{\mu }}\right)$ should satisfy

$$\psi \left(k_\mu \right){\longrightarrow 0 \atop {\scriptstyle K_\mu \rightarrow
\;\infty}}\ \,\ \sum\limits_{m,r,s,t}^{} \psi \left({{k}_{\mu }}\right)<\infty$$

\section{Induced representations of the discrete Poincar\'e groups}

Let ${ \cal P}_{ +}^{\uparrow } ={T}_{4}{\times}_{5}SO\left({3,1}\right)$ be the Poincar\'e group
restricted to the integral Lorentz transformations and discrete traslations on the lattice with
the group composition

\begin{equation}
\left({ a,\wedge }\right)\left({ a'\wedge '}\right) =\left({a+\wedge a',\wedge \wedge
'}\right)
\end{equation}

In order to construct irreducible representations we follow standard method 

(1) Choose an $UIR\ {D}^{\stackrel{o}{k}}\left({a}\right)$ of the translation group $T_4$

(2) Define a little group $H \in  SO(3,1)$ by the stability condition

\begin{equation}
h\in H\ \ :\ \ {D}^{\stackrel{o}{k}}\left({{h}^{-1}a}\right)={D}^{\stackrel{o}{k}}\left({a}\right)
\end{equation}

(3) This condition leads to the following decomposition of the $UIR$ of $T_4\times_sH$

\begin{equation}
{D}^{\stackrel{o}{k} ,\alpha }\left({a,h}\right)
={D}^{\stackrel{o}{k}}\left({a}\right)\otimes {D}^{\alpha }\left({h}\right)
\end{equation}

(4) Choose coset generators $c$ of $T_4\times_sH$ constructed from the group action

\begin{equation}
\left({\tilde{a},\tilde{\wedge }}\right)c=c'\left({a,h}\right)
\end{equation}

(5) Then the induced representations is: 

\begin{equation}
{D}^{\stackrel{o}{k},\alpha }\left({\tilde{a}\rm ,\tilde{\wedge
}}\right)={D}^{\stackrel{o}{k},\alpha }\left({a,h}\right)\delta
\left({{\left({c'}\right)}^{-1}\left({\tilde{a},\tilde{\wedge
}}\right)c,\left({a,h}\right)}\right)
\end{equation}
this is an $UIR$ of ${\cal P}_{+}^{\uparrow }$

In the case of the discrete Pincar\'e group for the massive case,
$\stackrel{o}{k}={m}_{0}c\left({1,0,0,0}\right)$ and the little group is the
cubic group that satisfies condition (81).

For the coset representative $c\equiv \left({0,\wedge }\right)$ we can choose the integral
Lorentz transformations $\wedge \equiv L\left({k}\right)$ that take $\stackrel{o}{k}$
into an arbitrary discrete momentum. The Dirac delta function in (84) is zero unless

\[\left({a,h}\right)
=\left({0,{L}^{-1}\left({k'}\right)}\right)\left({\tilde{a},\tilde{\wedge
}}\right)\left({0,L\left({k}\right)}\right)=\left({{L}^{-1}\left({k'}\right)\tilde{a},{L}^{-1}\left({k'}\right)\tilde{\wedge
}L\left({k}\right)}\right)\]

Substituting in (84) and using (82) we get 

\begin{equation}
{D}_{k'k}^{\stackrel{o}{k},\alpha }\left({\tilde{a}\rm ,\tilde{\wedge }}\right)
={D}_{k'k}^{\stackrel{o}{k}}\left({{L}^{-1}\left({k'}\right)\tilde{a}}\right){D}_{k'k}^{\alpha
}\left({{L}^{-1}\left({k'}\right)\tilde{\wedge }L\left({k}\right)}\right)
\end{equation}

The spinor representation of the second factor is given with respect to the element
${L}^{-1}\left({k'}\right)\tilde{\wedge }L\left({k}\right)$ that belongs to the little group,
$SU(2)$, and this group becomes irreducible when restricted to the spin 1/2 or spin 1
representation [12]. The first factor can be written: 

\begin{equation}
{D}_{k'k}^{\stackrel{o}{k}}\left({{L}^{-1}\left({k'}\right)\tilde{a}}\right)
={D}^{k'}\left({\tilde{a}}\right)
\end{equation}
where $k'={\left({{L}^{-1}\left({k'}\right)}\right)}^{T}\stackrel{o}{k}$ are all the points that defined
the $UIR$ of translation group and belong to the orbit.

We want to characterized the discrete orbits by functions in the momentum space that vanish on
the orbit points and only in these points. A natural way to construct these functions is through
the Dirac equation in momentum space (59). Multiplying this equation from the left by
$\left({{\gamma }_{\mu }{p}_{\mu }+{m}_{c}}\right)$ we get

\begin{equation}
\left({{p}_{\mu }{p}_{\mu }-{m}_{c}^{2}}\right)\psi
\left({p}\right)=0
\end{equation}

The $p_{\mu}$ have different physical meaning as we stressed in the realization of Dirac
equation in position space. We have to check whether the constraints (88) satisfy the following
conditions:

(1) they should be polynomials or infinite product of polynomials that vanish on the orbit points,

(2) the constraints should admit a periodic extension of the momentum space,

(3) the constraints should vanish on a orbit and they must be Lorentz invariant,

(4) the constraints should vanish only on the points of the orbit,

(5) when the lattice spacing goes to zero, the difference equation should go to the continuous
Minkowski limit.

In the Fourier transform of type I and II the $p$ variable are related to the physical momentum
$k$ by the expresion

$${p}_{\mu }={\frac{2}{\varepsilon }}tg{\frac{\pi }{N}}{m}_{\mu }\ \ \ {m}_{\mu
}=0,1,\ldots N-1$$

or
$${p}_{\mu }={\frac{2}{\varepsilon }}tg\ \pi \varepsilon {k}_{\mu }\ \ \ {k}_{\mu }\in
{\mathcal{Z}}$$

Due to the special trigonometric function this expression is periodic in momentum space.
Nevertheless when Lorentz transformations are applied to the components of physical momentum the
new momentum do not satisfies the constraint equation and the contraints vanish at different
points. Therefore conditions (3) and (4) are violated, but they can be recovered in the
asymptotic limit when $\varepsilon \rightarrow 0$

$$\left({\left({\pi \varepsilon {k}_{\mu }}\right)\left({\pi \varepsilon {k}_{\mu
}}\right)-{m}_{0}^{2}}\right)\psi \left({{k}_{\mu }}\right)=0$$

In the case of Fourier transform of type III the $p_{\mu}$ coincide whit the physical momentum
whose discrete values are the points of the orbit given by ${\wedge }_{\mu}{\stackrel{o}{p}}_{\mu}$ and
therefore condition (3) and (4) are fullfiled, although (2) is violated. In all the three cases
the wave equation on position space, as described in section 4 gives in the asymptotic limit,
the continuous Dirac equation.

\section{Concluding remarks}

We have attempted a new program for introducing Poincar\'e symmetry on the lattice, that was
considered broken, as many authors have claimed [13]. So far some points of this program have
been achieved; such as, realization of all the integral Lorentz transformation and is
representation in 2-dimensional space. Several $UIR$ of the translation group on the lattice,
some versions of the Fourier transform with discrete position and discrete or continuous
momentum, the Dirac equation on the lattice in position space. As far as the induced
representation of the Poincar\'e group, following similar technic as in the continuous case,
leads to some inconsistensy: for the type I and II of the $UIR$ of translation group the scheme
is irreducible and invariant only in the asymptotic limit, but they satisfy all the requirement
for the induced representation. For the type III the representation is Lorentz invariant and
irreducible, but the requirement of the orbit condition necessary for the induced representations
is kept only in the asymptotic limit.

\vskip 1truecm  \noindent  {\Large \bf Acknowledgments}
\vskip 0.5cm

One of the authors want to expressed his gratitude to the Director of the Institut
f\"{u}r  theoretische Physik, T. Universit\"{a}t Tubingen, where part of this
work was done, for the hospitality. This work has been partially supported by D.G.I.C.Y.T. contract
\#Pb94-1438 (Spain).


\begin{thebibliography}{00}

\bibitem{} A. Schild, ``Discrete space-time and integral Lorentz transformation'', Can. J. Math.
{\bf 1}, 29 (1948). Appendix. 

\bibitem{} V. Kac, {\it Infinite dimensional Lie Algebras}, Cambridge U. Press (1991)pp. 69-71.

\bibitem{} M. Lorente, ``Cayley parametrization of semisimple Lie groups and its application to
Physical Laws in a (3+1)-dimensional cubic lattice'', Int. J. Theor. Phys. {\bf 11}, 213-247
(1974).

\bibitem{} C. M{\o}ller, {\it The theory of Relativity}, Oxford Clarendon Press, 1952, p. 42.

\bibitem{} Ref. 3, p. 221.

\bibitem{} M. Lorente, ``A new scheme for the Klein-Gordon and Dirac fields on the lattice with
Axial Anomaly'', {\it J. Group Th. in Phys.}. {\bf 1} 105-121 (1993), p. 107.

\bibitem{} See M. Crentz, {\it Quarks, gluons and lattices}, Cambridge U. Press, 1983, p. 15. See
also, I. Montvay, G. M\"{u}nster {\it Quantum Field on the lattice}, Cambridge U. Press, 1994.

\bibitem{} M. Lorente, ``Discrete Reflection Groups and Induced Representations of Poincar\'e
Group on the Lattice'' {\it Symmetries in Science IX} (B. Gruber) Academic Press (1997).

\bibitem{} P. Kramer, The Lorentz group and Dirac equation (?)

\bibitem{} L. Fonda, G.C.Ghirardi, {\it Symmetry Principles in Quantum Physics}, Marcel Dekker
1970, p. 309.

\bibitem{} U.H. Niederer, L.O'Raifertaigh, ``Realization of the Unitary Representations of the
Inhomogeneous Space-time groups'', I and II, Forsch. Phys {\bf 22}, 111-129, 131-157 (1974).

\bibitem{} Melvin Lax, Symmetry principles in Solid State and Molecular Physics, John Wiley \& sons, New York
1974, p. 431-2, 436-8.

\bibitem{} I. Montvay, ``Supersymetric gauge theories on the lattice'', Lattice 96, Nuclear
Physics B (Proc. Suppl.) {\bf 53} (1967), 853-5.

\bibitem{} P. Kramer, M. Lorente, ``Discrete and continuous symmetry via, induction and duality'',
Proceedings Symmetries in Science X (this volume).

\end{thebibliography}
\end{document}